\documentclass[showpacs,pra,twocolumn,aps]{revtex4}
\usepackage{graphicx}
\usepackage{psfrag}
\usepackage{bm}
\begin{document}
\title{Zero temperature damping of 
Bose-Einstein condensate oscillations by vortex-antivortex pair creation}
\author{Petr O. Fedichev$^{1,2}$, Uwe R. Fischer$^1$, and Alessio Recati$^{1,3}$}
\affiliation{$^{1}$Leopold-Franzens-Universit\"{a}t Innsbruck, 
Institut f\"{u}r Theoretische Physik, Technikerstrasse 25, 
A-6020 Innsbruck, Austria \\
$^2$Russian Research Center Kurchatov Institute, Kurchatov Square, 123182 Moscow, Russia \\ 
$^3$Dipartimento di Fisica, Universit\`a di Trento and BEC-INFM, 
I-38050 Povo, Italy}

\begin{abstract} 
We investigate vortex-antivortex pair creation in a supersonically expanding 
and contracting quasi-2D Bose-Einstein condensate at zero temperature. 
For sufficiently large amplitude condensate oscillations, pair 
production provides the leading dissipation mechanism. 
The condensate oscillations decay in a nonexponential fashion,  
and the dissipation rate depends strongly on the oscillation amplitude. 
These features allow to distinguish the decay due to 
pair creation from other possible damping mechanisms.  
Experimental observation of the predicted oscillation behavior
of the superfluid gas provides a direct confirmation of the 
hydrodynamical analogy of quantum electrodynamics
and quantum vortex dynamics in two spatial dimensions. 
\end{abstract}
\maketitle

The process of electron-positron 
pair creation is well-established in quantum electrodynamics 
since the seminal work of J. Schwinger \cite{Schwinger}. Later on, it became apparent 
that the hydrodynamics of vortices in two-dimensional (2D) superfluids can
be mapped onto 2+1D  electrodynamics with vortices playing the role of charged particles,
and phonons the role of photons \cite{Popov}. In this analogy, the superfluid density and the supercurrent  are 
acting as the magnetic and electric fields on the vortices whose circulation is the charge. 
The Schwinger vacuum breakdown is a phenomenon occuring whenever the electric field exceeds the 
magnetic field (in cgs units), which corresponds in the analogy to the instability of a supersonic flow 
with respect to the spontaneous creation of vortex-antivortex pairs from the superfluid vacuum.

Vortices in Bose-Einstein condensates have been observed and studied experimentally intensely in the last couple of years, e.g., in 
\cite{madison,VortexRaman,Ketterle,Cornell,Rosenbusch}. 
Here, we suggest an experiment in a quasi-2D Bose-condensed gas revealing the existence of irreversible 
condensate dynamics at zero 
temperature as the result of the Schwinger pair creation instability. 
To argue that vortex-antivortex pair creation is 
the dominant source of dissipation, we use the fact that a quasi-2D
BEC in a time-dependent harmonic trap 
has a peculiar feature: There is a time dependent transformation (the 
so-called "scaling transformation"),   
using which  the problem can be exactly solved, 
because the time dependence can effectively be removed 
from the Hamiltonian \cite{Scaling,Castin,Kagan}. 
This scaling property also holds for linearized equations describing the evolution 
of small density and phase perturbations (Bogoliubov quasiparticle excitations), 
propagating on top of the moving superfluid. Therefore, initial perturbations cannot grow, 
and the instability mechanisms known from classical hydrodynamics play no role 
\cite{LLFluid}. This stability against perturbations implies that at very low temperatures,  
condensate oscillations are practically undamped 
(it has been measured that the quality factor 
$Q\gtrsim 2000$ \cite{DalibardBreathing}).

On the other hand, the superfluid velocity in the scaling solution grows linearly towards the condensate
border, while the local density decreases. Therefore the outer region of the cloud is always supersonic and, 
according to the Landau criterion for superfluidity, can host instabilities.
Vortices are {nonlinear} excitations above the superfluid ground state, 
so that they are not protected by the scaling symmetry, which holds for 
{linear} excitations. 
Spontaneous vortex-antivortex pair creation, 
analogous to the Schwinger process, is an {\em intrinsic} 
instability mechanism, and constitutes a source of dissipation 
already at zero temperature, without any need for 
a symmetry breaking external perturbation.

In the following, we explicitly analyze the Schwinger instability of a supersonically expanding 
and contracting BEC in a time dependent quasi-2D harmonic trap. 
We show that for sufficiently large condensate oscillations vortex-antivortex
pair production provides the dominant dissipation mechanism. 
Furthermore, the condensate oscillations decay in a nonexponential fashion and the dissipation rate depends strongly on the oscillation amplitude. These features allow one to distinguish
experimentally the decay due to pair creation from the previously studied damping mechanisms. We note that the suggested
zero temperature 
damping mechanism is intrinsically different from that discussed in \cite{Kagan}, where the dissipation is due to the energy 
transfer from the radial condensate motion to the longitudinal modes in an elongated cylindrically-symmetric condensate. 
This mechanism can only work if the condensate is sufficiently long, whereas we confine ourselves to the case of a quasi-2D sample, 
for which any motion along the $z$-axis is suppressed. 

The analogy of 2D vortex dynamics 
with electrodynamics is most easily established by noting that 
the expression for the 2D Magnus force 
${\bm F}_M = 2\pi \rho {\bm e}_z \times ( \dot {\bm X} -{\bm v}_s)$ leads to the identification
of  ${\bm E}=\rho {\bm v}_s\times {\bm e}_z$ and ${\bm B}=-\rho {\bm e}_z$ 
with the ``electric'' and ``magnetic'' fields, by comparing 
with the Lorentz force ${\bm F}_L = q   ({\bm E} + \dot {\bm X} 
\times {\bm B})$.  Here, $\dot {\bm X}$ and ${\bm v}_s$ are vortex and local superflow
velocities, respectively, and $\rho$ is the local density.  
The circulation ($2\pi$ in our units with $\hbar=m=1$) 
is the ``charge'' $q$ (cf., e.g., \cite{donnelly1,arovas,annalspaper}). 
The self-energy of a (widely separated) single vortex pair 
is $2E^0_v=2\pi\rho\Lambda$, 
with $\Lambda=\ln(R/a_c)$, where $R$ is the size of the pair.  
We will use in this expression for the pair
energy that the vortex core size in a dilute superfluid is given by  
$a_c=1/c_s$, where $c_s$ is the speed of sound.  
The inertial rest mass of a vortex stemming from compressibility,  
$m_v = E^0_v/c_s^2$, is (for large condensates) 
of ``electrodynamical'' origin:  
It stems from the self-interaction of a moving vortex with the long-range 
flow and density fields it induces inside the surrounding superfluid medium. 
Since the ``electromagnetic'' fields (the density and velocity perturbations) represent 
``relativistic'' particles (phonons), the vortex mass diverges if the velocity of the 
accelerated vortex approaches the speed of sound, in the same manner in which the mass of a charged 
ultrarelativistic particle diverges in conventional electrodynamics. 
We assume in what follows 
that other possible contributions to the vortex mass (see, e.g., 
the backflow mass contribution discussed in \cite{baymchandler})
remain regular if the local superfluid velocity approaches the speed of sound.
These contributions are therefore subdominant 
for ``relativistically'' moving vortices.

\begin{figure}[hbt]
\includegraphics[width=8.5cm]{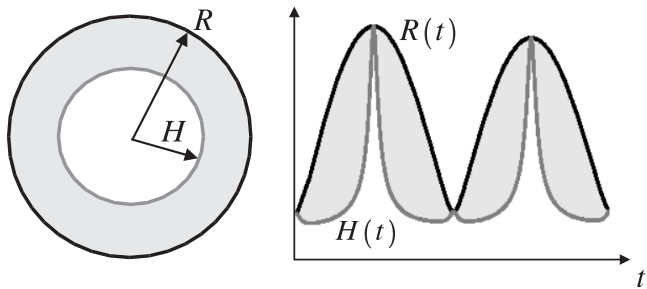}
\caption{\label{Fig1} An oscillating, cylindrically symmetric
 quasi-2D condensate. 
The shaded region designates the region of space in which 
the speed of sound is exceeded by the oscillating condensate,  
and vortex pair creation takes place; $H=H(t)$ is the horizon location and 
$R=R(t)$ the Thomas-Fermi radius of the condensate.}
\end{figure}

We consider a quasi-2D superfluid Bose gas in a time-dependent 
isotropic harmonic trapping 
potential $V({\bm x},t)=\frac12\omega^2(t) (x^2+y^2)$, with ${\bm x}=(x,y)$. 
It is a well-established fact that the hydrodynamic solution for 
density and velocity of motion in a harmonic potential 
with arbitrary time dependence may be obtained 
from a given initial solution by a scaling procedure \cite{Scaling,Castin}.  
Defining the scaled coordinate vector ${\bm r}_b={\bm x}/b$, 
the rescaled density and velocity are given by 
\begin{eqnarray}
\rho({\bm x},t)&=&\frac{1}{b^2}\sigma({\bm r}_b)=\frac{\rho_0}{b^2}
\left(1-\frac{r_b^2}{R_0^2}\right),\\
{\bm v}_s({\bm x},t)&=&\frac{\dot b}{b}{\bm x}\,.
\end{eqnarray}
Here, we assume the superfluid to be described initially 
within the Thomas-Fermi (TF) approximation (that is, the condensate is large enough to neglect the quantum 
pressure);  $\rho_0$ is the initial central density and $R_0$ the initial TF radius, 
such that  $R= R(t)= b(t) R_0$ is the instantaneous TF radius of the cloud. 
The energy functional has  in the TF approximation the form
\begin{equation}
{\cal E}(b,\dot b)=\frac{1}{2 b^2}\int d^2{\bm r}_b\left[\left(\omega^2+
\frac{\dot b ^2}{b^2}\right)b^4 r_b^2\sigma+g\sigma^2\right],
\label{energy-func}
\end{equation}
with $g$ the interaction strength, which depends in the present quasi-2D case on the tight confinement 
in $z$-direction and on the density of the condensate \cite{Petrov}.
This leads to an effective Hamiltonian for the dynamical variable $b$,  
\begin{equation}
{\cal E}(b,\dot b)=\left(\frac{\alpha}{2}\dot
b^2+\frac{\alpha}{2}\omega^2 (t) b^2+\frac{\beta}{2 b^2}\right),
\label{energy-b}
\end{equation}
where $\alpha=
\pi \rho_0 R_0^4/6$ and 
$\beta= 
\pi \rho_0^2 g R_0^2/3$. 

Consider a situation in which the external trap frequency is changed 
from $\omega_{\rm in}$ to $\omega_f \ll  \omega_{\rm in}$, on a time 
scale much less than the inverse initial trap frequency.
As a consequence, the gas undergoes a large amplitude monopole oscillation
with frequency $2\omega_f$ \cite{PitaRosch}. 
At sufficiently low temperatures (below the Kosterlitz-Thouless temperature), 
the initial state of the superfluid contains bound vortex-antivortex pairs, 
i.e.\/\,\,\,topological excitations, 
which can be unbound by the action of the Magnus force in the
(time dependent) supersonic flow region.  
Indeed, for an oscillating condensate, there exists a region, the border 
of which is called horizon (cf. Fig.\,\ref{Fig1}), where the 
superfluid velocity magnitude  
$v_s$ is larger than the local sound velocity $c_s=\sqrt{g\rho}$. 
The speed of sound is exceeded at  
the {\em horizon radius} 
\begin{equation}
H(t)=\frac{b(t) R_0}{\sqrt{\gamma^2(t)+1}}\,,
\end{equation} 
where $\gamma 
= \sqrt{2}\dot b b/\omega_{\rm in} $.

Beyond the horizon, the vortices and antivortices get accelerated during condensate 
evolution and separate at local superflow velocities larger than that of sound. 
This is analogous to the Schwinger pair creation process in quantum electrodynamics.
It is important to recognize 
that the flow we consider is inhomogeneous  and time dependent by default.  
Consequently, the argument that there is no pair creation possible  
because one could always use the underlying Galilean invariance to 
``transform away'' the background flow, 
does not apply to our situation.

In a simple model of the 2+1D vacuum pair creation 
instability, which exploits directly the analogy to Schwinger pair creation 
in quantum electrodynamics, 
the pair production rate  $\Gamma$ per unit area can be written as \cite{italians}
\begin{equation}
{\Gamma} =\frac{1}{4\pi^2
c_s^2} {\cal F}^{3/4}\sum_{n=1}^\infty\frac{(-1)^{n+1}}{n^{3/2}}
\exp{\left(-\frac{\pi n {(E_v^0)}^2}{\sqrt{\cal F}}\right)},
\label{it-formula}
\end{equation} 
where we have defined ${\cal F} = E^2c_s^2-B^2c_s^4$
and set, within logarithmic accuracy, the vortex pair size in $E_v^0$
equal to the Thomas-Fermi radius of the condensate.  
The above relation holds  for locally supersonic motion, i.e.,  
if $|{\bm E}|/|{\bm B}|> c_s$ (${\cal F}> 0$). The value of the 
prefactor in front of the exponential in the above expression 
is subject to changes which are due to the microscopic details 
of vortex motion. 
We display its value, stemming from taking literal 
the analogy to quantum electrodynamics also on the level of quantum 
fluctuations (to one loop order), 
for numerical concreteness. However, the behavior 
of $\Gamma$ for $|{\bm E}|/|{\bm B}|\gtrsim c_s$ is dominated by 
the hydrodynamical exponent, 
whose value is independent of microscopic physics, and 
specifically by the $n=1$ term in the above sum.

Assuming that the vortex density is low, the 
energy dissipation rate, $\dot\epsilon$, is obtained by multiplying 
Eq.\,(\ref{it-formula})  by the rest energy of the widely separated vortices  
$2E_v^0$, and integrating over the area of the TF domain. This results in  
\begin{equation}
\dot\epsilon=\frac{g^{1/2}\Lambda\rho_0^3 R_0^2 }{2 b^4}
\frac{\gamma^{8}}{(\gamma^2+1)^{13/4}}
F \left(\frac{\Lambda^2}{g}\frac{1}{\sqrt{\gamma^2+1}}\right), 
\label{epsdot}
\end{equation}
where we introduced the function 
\begin{equation}
F (\lambda )= \sum_n
\frac{(-1)^{n+1}}{n^{3/2}}
\int_0^1 d\eta\, \eta^{\frac34}(1-\eta)^{\frac94}
e^{-\pi n\lambda
\sqrt{{(1-\eta)}/{\eta}}}.
\end{equation}
Since $\gamma$ is proportional to $\dot b b$, the Schwinger dissipation rate (\ref{epsdot}) can give rise to a
measurable effect only if the condensate oscillation amplitude is sufficiently large, which 
implies $\omega_f \ll \omega_{\rm in}$. 
In order to provide some analytical results, we consider a simple quasistationary perturbation 
theory approach.  Consequently, we assume that the dissipation rate is small and therefore that the  
energy of the system in Eq.\,(\ref{energy-func}) is a slowly varying 
function within each oscillation period. Then the equation of motion for the scaling 
parameter $b$ can be found from 
\begin{equation}
\frac{d}{dt}{\cal E} (b, \dot b)=-\dot\epsilon \,. \label{ThisEq}
\end{equation}
In the absence of dissipation ($\dot \epsilon=0$), 
the range of $b$  is between $b_{\rm min}=1$ and $b_{\rm max} = \omega_{\rm in}/\omega_f$. Since 
$b_{\rm min}=1 \ll b_{\rm max}$, we can approximately set $b_{\rm min}\simeq 0$. 
One can then write $\gamma^2=2(\omega_f^2/\omega_{\rm in}^2)b^2
(b_{\rm max}^2-b^2)$. 
In a dilute gas, in the TF limit, 
the argument of $F$ is large, $\Lambda^2\gg g \sqrt{\gamma^2+1}$, and 
the dynamical equation (\ref{ThisEq}) for $b$ takes the simpler form: 
\begin{equation}
\ddot b + \omega_f^2 b - \frac{\omega_{\rm in}^2}{b^3}= 
-\frac{ {\mathcal D}}{\omega_{\rm in}^5 }{\dot b}^7b^4\,,
\label{dyn-b}
\end{equation}
where the constant 
\begin{equation}
{\mathcal D}=\frac{48}{\pi^8}\frac{g^{7/4}\sqrt{gN}}
{\left(\ln\left[4\sqrt{{gN}/{\pi}}\right]\right)^{\frac{11}2}}
\sum_n\frac{(-1)^{n+1}}
{n^8}.\end{equation}
Using the equation (\ref{dyn-b}),  
$b\dot b$ can be expressed in terms of $b$ only,  
$b^2\dot b^2=\omega_f^2 b^2(b_{\rm max}^2-b^2)$. 
The oscillation energy lost in a period is then given by
\begin{equation}
I_E=\frac{7\pi}{1024}{\mathcal D}\frac{\omega_f^7}{\omega^5_{\rm in}}b_{\rm max}^{12}\,.
\end{equation}
The energy decrease 
rate for $b_{\rm max}$ is obtained from the equation
\begin{equation}
\frac{d}{dt}{\cal E} (b_{\rm max})      
=\frac{\omega_f}{\pi} I_E.
\end{equation}
Thus one obtains for the oscillation peak value the following expression   
\begin{equation}
b_{\rm max}(t)=\frac{b_{\rm max}(0)}{(1+ {\mathcal D'}b_{\rm max}^{10}(0)\omega_f t
)^{1/10}}, \label{bmax}
\end{equation}
where ${\mathcal D'}=\frac{35}{512}
(\omega_f/\omega_{\rm in})^5{\mathcal D}$.
Our perturbation theory approach is valid as long as $b_{\rm max}^{10}(0){\mathcal D'}\ll 1$.
\vspace*{-2em}
\begin{center}
\begin{figure}[b]
\psfrag{X}{\LARGE $\!\!\!\!\!\frac R{R_0}$}
\hspace*{3.3em}
\includegraphics[width=9.5cm]{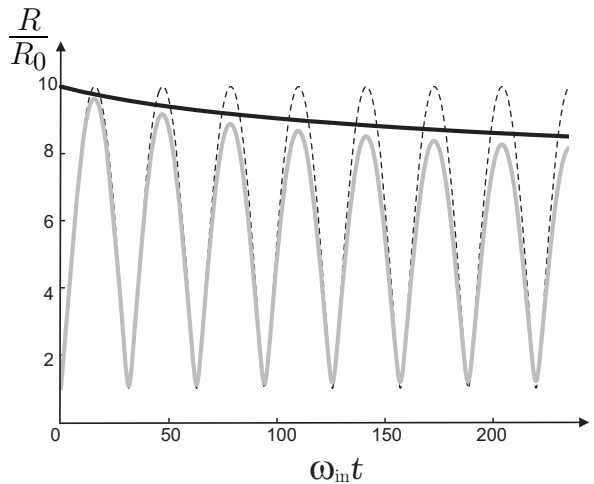}
\caption{\label{Fig2} Damping of condensate oscillations due to 
vortex-antivortex pair creation, with $N=10^4$, $g=1$, and $\omega_f = 0.1\, \omega_{\rm in}$, where  the radius $R$ is in units of the original 
Thomas-Fermi size $R_0$.
The black solid line is the envelope $b_{\rm max}$ from Eq.\,(\ref{bmax}).
The grey solid line is the damped breathing
mode oscillation obtained from numerically solving Eq.\,(\ref{dyn-b}). 
For comparison, the dashed line represents
the oscillation of the superfluid gas without pair creation taking place.}
\end{figure}
\end{center}

The damping of condensate oscillations due to vortex-antivortex pair creation is represented
in  Fig.\,\ref{Fig2}, where we show the numerical solution of
the dynamical equation Eq.\,(\ref{dyn-b}) (grey solid line), the approximate solution for the peak 
amplitude (black solid line), and  for comparison the free oscillation without pair creation (dashed line).
The parameters used in the numerical integration for the plot are $N=10^4,\;g=1$, and  
for the final trapping frequency $\omega_f=0.1\,\omega_{\rm in}$. 
These parameters are consistent with the argument of $F$ being large,
$\Lambda^2\gg g \sqrt{\gamma^2+1}$, so that 
Eqs.\,(\ref{dyn-b})--(\ref{bmax}) hold. 
The envelope $b_{\rm max}(t)$ is seen to decay 
very slowly and in a nonexponential fashion, 
governed by the TF exponent $\frac1{10}$ in Eq.\,(\ref{bmax}).
For realistic parameters, we conclude from Fig.\,\ref{Fig2} that
an observable damping effect for the condensate oscillations is obtained.

The scaling parameter evolution can be described by Eq.\,(\ref{ThisEq}) only for 
sufficiently short times, when the total density of vortices produced is still low.
At later times, the vortex-antivortex plasma can decrease the superfluid current in the same way
as the electron-positron plasma can screen the electric field.  This is an interesting collective 
effect, which however requires a more elaborate treatment. 

We described an intrinsic damping mechanism for large amplitude condensate 
oscillations in a quasi-2D 
Bose gas at zero temperature. 
The dissipation originates from spontaneous creation of 
vortex-antivortex pairs and depends on the shape 
and dynamics of the supersonic flow region. 
The results we presented therefore depend strongly on the oscillation amplitude. This feature can 
be used to distinguish the effects of pair production from other 
possible dissipation mechanisms. The scaling solution not only exists for
the discussed monopole modes, but also for quadrupole oscillations, so that, e.g., effects resulting
from a rotating superfluid on the pair creation process may be studied.    
Observation of the predicted oscillation behavior of the superfluid gas provides direct confirmation 
of the hydrodynamical analogy of quantum electrodynamics and quantum vortex dynamics in two spatial 
dimensions, and would put this analogy to its first real experimental test. 
Such confirmation would, then, give further motivation to the program 
of studying analogies between high energy physics, cosmology and condensed matter systems 
\cite{GrishaPhysicsReports}.  

The outlined mechanism for dissipation is not confined to quasi-2D samples. 
In strongly elongated 3D condensates, 
the scaling solution also applies, and the vorticity is generated in the form of vortex rings, with the total
vorticity integrated over the sample volume still zero. However, as already mentioned above, 
for a 3D condensate the effect of vortex ring creation can be masked by possibly stronger damping mechanisms, like the parametric resonance 
discussed in \cite{Kagan}.

We thank L.\,P. Pitaevski\v\i\/ and G.\,E. Volovik 
for critical remarks and helpful comments on the manuscript, and P. Zoller
for discussions. P.\,O.\,F. 
has been supported by the Austrian Science Foundation FWF and the Russian Foundation for Basic
Research, U.\,R.\,F. by the FWF, and A.\,R. by the  European Union under Contract No. HPRN-CT-2000-00125.

\end{document}